\documentclass[10pt,conference]{IEEEtran}
\IEEEoverridecommandlockouts

\usepackage{amsmath,amssymb,amsfonts}
\usepackage{algorithmic}
\usepackage{graphicx}
\usepackage{textcomp}
\usepackage{wrapfig}
\usepackage{xcolor}
\usepackage[hidelinks]{hyperref}

\usepackage{pbalance}

\makeatletter
\newcommand{\linebreakand}{%
  \end{@IEEEauthorhalign}
  \hfill\mbox{}\par
  \mbox{}\hfill\begin{@IEEEauthorhalign}
}
\makeatother

\usepackage{listings}
\usepackage{fancyvrb}
\usepackage{framed}
\usepackage[listings,skins]{tcolorbox}
\usepackage[skipbelow=\topskip,skipabove=\topskip]{mdframed}
\mdfsetup{roundcorner=0}

\begin{document}
\title{Towards Safety and Security Testing of Cyberphysical Power Systems by Shape Validation
}

\author{
	\IEEEauthorblockN{
        Alexander Geiger\IEEEauthorrefmark{1},
        Immanuel Hacker\IEEEauthorrefmark{1}$^{,}$\IEEEauthorrefmark{2},
        Ömer Sen \IEEEauthorrefmark{1}$^{,}$\IEEEauthorrefmark{2} and
        Andreas Ulbig\IEEEauthorrefmark{1}$^{,}$\IEEEauthorrefmark{2}
        }
	
	\IEEEauthorblockA{
        \IEEEauthorrefmark{1}Digital Energy at Fraunhofer FIT, Aachen, Germany
        \\Email: alexander.geiger@fit.fraunhofer.de
        }
        
    \IEEEauthorblockA{
        \IEEEauthorrefmark{2}IAEW at RWTH Aachen University, Aachen, Germany
        }
 }
 
\maketitle

\maketitle

\begin{abstract}
The increasing complexity of cyberphysical power systems leads to larger attack surfaces to be exploited by malicious actors and a higher risk of faults through misconfiguration. We propose to meet those risks with a declarative approach to describe cyberphysical power systems and to automatically evaluate security and safety controls. We leverage Semantic Web technologies as a well-standardized framework, providing languages to specify ontologies, rules and shape constraints. We model infrastructure through an ontology which combines external ontologies, architecture and data models for sufficient expressivity and interoperability with external systems. The ontology can enrich itself through rules defined in SPARQL, allowing for the inference of knowledge that is not explicitly stated. Through the evaluation of SHACL shape constraints we can then validate the data and verify safety and security constraints. We demonstrate this concept with two use cases and illustrate how this solution can be developed further in a community-driven fashion.
\end{abstract}


\section{Introduction}
 \subsection{Motivation}
As energy grids increasingly rely on communication infrastructure for their operation, they exhibit an larger attack surface towards malicious actors. Therefore, cybersecurity plays an ever increasing role in the daily business of power system operators. Furthermore, risk of faults through misconfiguration increases in more complex cyberphysical power systems, potentially compromising the (fail) safety of the infrastructure.
Incidents such as the attack on the Ukranian power grid in 2016 \cite{Whitehead2017-eo}, as well as the significant frequency deviation in the continental transmission grid in 2022 \cite{ENTSO-E2019-oo} are first examples of those risks. The transition to a smart grid will only increase the likelihood of such incidents, if effective countermeasures are not put in place. Modern risk management is still dominated by manual audits checking security controls, we strive to complement this practice with more rigorous and automated reasoning, reducing the risk of human error.

\subsection{Related Work}
In this paper we present a declarative approach to safety and security testing of cyberphysical power systems that utilizes technologies from the Semantic Web group of standards. Similar approaches have been presented in past work, however none covers all aspects of our proposed approach.
The centerpiece of our application is a semantically rich data model and may be considered to be a knowledge graph. Using knowledge graphs to manage cybersecurity information has been attempted by many projects. \cite{Sikos2023-bz} provides a broad literature review of cybersecurity knowledge graphs and \cite{Hollerer2024-xj} contains a systematic mapping study of ontologies considering safety, security and operation aspects in operational technology (OT) in cyber-physical systems. The latter survey identifies the need for more holistic modeling for effective risk management.
The project Unified Cybersecurity Ontology (UCO) \cite{Syed2016-mv} is a notable effort to model cybersecurity concepts in an ontology. It provides an open-source ontology of useful cybersecurity terminology and is led by the MITRE corporation. At it current state however it lacks detail and capacity for reasoning. An integration of this ontology in our application may be considered in the future. Also in the power systems domain, many interesting modeling efforts are ongoing. \cite{Abdelmalak2022-py} provides a useful overview regarding recent approaches. Most similar to our effort, by the combination of multiple models of varying degree of abstraction is \cite{Vereno2024-il}. The focus of this work however is more on dynamic analysis (system-of-systems co-simulation) instead of static rules-based analysis.
More dynamic approaches to safety and security testing are often connected with the concept of Digital Twins, e.g. \cite{Atalay2020-eh} or \cite{Marksteiner2021-tp}. This usually implies simulation or emulation of the physical system. This approach has the merit of finding potential dynamic vulnerabilities that could not have foreseen. In our work, we focus on static analysis of architecture.
Two initiatives very close to ours are the Semantic Inference Model for Security in Cyber-Physical Systems using Ontologies (SIMON) \cite{Venkata2019-ad} and the Ontology-driven approach for Cyber-Physical Security Requirements meta-modelling and reasoning (Onto-CARMEN) \cite{Blanco2023-sv}.
Both initiatives provide security testing through deductive reasoning on semantic models of cyber-physical systems. However, while SIMON includes knowledge from third parties, the system itself is not open and can therefore not be extended externally. Neither is Onto-CARMEN, for which the decision was made to design an ontology from scratch. While the technologies chosen for SIMON are not known, Onto-Carmen too chooses Semantic Web technologies. However, they are yet to adopt SHACL to validate security constraints. \cite{Larhrib2024-mn} developed a similar approach to validate semantic data against constraints. The authors identify the application of their approach to the security domain as potential future work.

\subsection{Contribution and Structure}
We propose to extend and combine existing data models for cyber-physical power systems, building on the approach presented in \cite{Hacker2025-xg}. We pair a declarative reasoning system with the model, so that through the application of well-defined rules, known risks can be eliminated. 
We leverage standards from the semantic web, namely RDF(S) , SPARQL and SHACL. We demonstrate two use cases for this approach, which indicate that the two incidents mentioned above could have been prevented with our proposed application. 
This approach is providing an additional use case to the model described in \cite{Hacker2025-xg} and it is reasonable to assume that there are more. We work on enabling open contribution by a larger community from academia and industry to this model. Through implementing their own use cases using the model, the community will have an incentive to extend the model further. This paper presents our vision in six sections. In the remainder of this section we present the context of our work in published literature. The following three sections describe our approach in technical detail, beginning with an architectural view and then focusing on modelling and reasoning. Section \ref{sec:case-studies} shows two use cases in each of which we implemented a test for a security or safety control as prototypes for the proposed application.
We conclude summarizing our key points and illustrate potential paths forward for the development of an implementation applicable to testing real infrastructure. 

\section{Target Application Architecture}
\label{sec:arch}
\begin{figure}
    \includegraphics[width=0.5\textwidth]{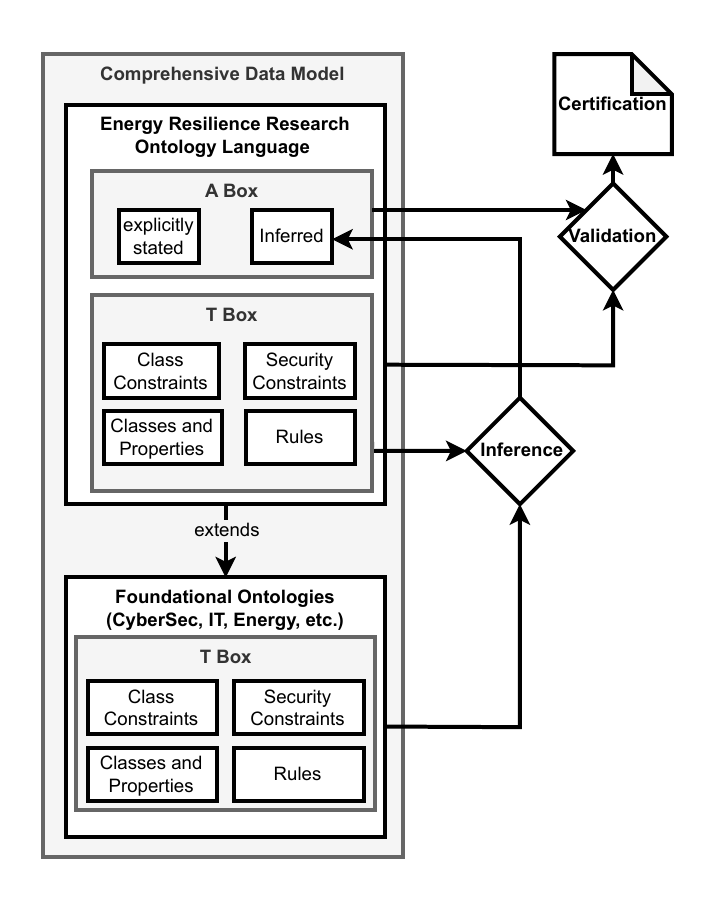}
    \caption{Data Model and Reasoning Steps}
    \label{fig:arch}
\end{figure}

Figure \ref{fig:arch} gives an overview of the architecture of our application.
It primarily consists of a comprehensive data model, drawing from our own ontology and external ontologies describing foundational concepts relevant to the domain of cyberphysical power systems. The data model defines the behavior of the application entirely, following a declarative programming style.
We utillize technologies from the Semantic Web set of standards for the sake of interoperability. The accessibility of these standards has led to a good variety of implementations of the standardized technologies and we will also work towards enabling third parties to extend the application. Furthermore, some relevant data is already available in RDF, such as the Common Information Model for electrical networks.
The ontologies define and describe concepts (classes and properties), using RDFS to provide a class hierarchy and SPARQL and SHACL to provide further semantics to the model. We describe in section \ref{sec:modeling} how we structure concepts used in our ontology. As indicated in figure \ref{fig:arch}, the model contains terminological (T Box) knowledge and assertional (A Box) knowledge. Assertional knowledge is that of concrete scenarios that can be tested using the terminological knowledge provided by or own and external ontologies. Besides stating the concepts, the T Box contains rules to derive additional knowledge from A Box statements. A simple reasoning mechanism using SPARQL is described in section \ref{sec:reasoning}.
Further, the T Box also containts two types of shape constraints: to test the completeness of the assertional knowledge (class constraints) and to test the adherence to security or safety controls (security constraints). 

\subsection{Further Use Cases}
This abstract architecture is not specific to the safety and security testing of cyberphysical power systems. We strive to provide a general framework for static analysis (through the reasoning over the data model through rule based extension and shape validation) of cyberphysical power system infrastructure, across abstraction levels and automation zones. Apart from security and safety applications as shown in this work, this approach may also underpin planning and economic analysis.
The data model underpinning this approach may also be transformed into configuration for simulations or real hardware, allowing for dynamic analysis. \cite{Hacker2025-xg} has used a previous iteration of this system for a simulative impact analysis of cyberattacks on behind-the-meter infrastructure. Practically any use case of or simulative analysis can theoretically be supported through this architecture. We will in future works explore which use cases this architecture suit more and which less. We will also work towards opening the framework to the wider research community, to facilitate the exploration of a variety of use cases. This will hopefully create synergies, especially through a more complete data model. 

\section{Modeling Approach}
\label{sec:modeling}
As described in the previous section, our modelling effort is not only aimed at supporting the use case presented in this paper, but to provide a resource for other analysis of cyperphysical power systems as well.
The concepts of our model are placed in a RDFS ontology, which allows for extension of the data model. SHACL shape constraints provide means to check data consistency.

The model underpinning our approach combines concepts from other models and creates relationships between them. We extend the approach for dynamic resilience assessment described in \cite{Hacker2025-xg} and subdivide the concepts along layers of abstraction. We include existing data models through import in case those are already available in a RDF-based format or manually transferring concepts from other formats. 

\begin{figure}
    \centering
    \includegraphics[width=0.5\textwidth]{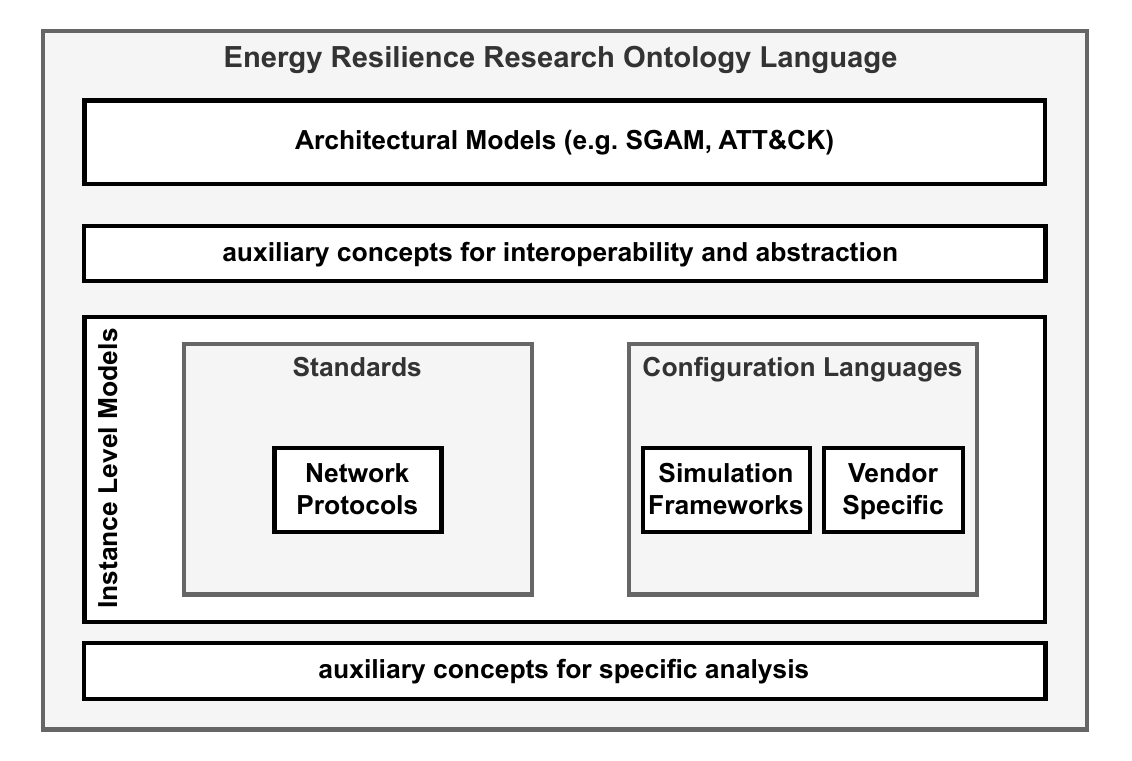}
    \caption{Division of concepts in the Energy Resilience Research Ontology Language}
    \label{fig:errol-structure}
\end{figure}
\subsection{Integration of Existing Models}
A wide variety of data models exist in the domain of cyberphysical power systems. We distinguish between architectural models, which describe architectures on different levels of abstractions and instance level models, which provide languages to describe concrete realisations of cyberphysical infrastructure. One example for an architectural model is SGAM, which currently is the architectural model we use to map infrastructure into an abstract 3D space spanned by the domain dimension (from bulk generation to behind-the-meter consumption and generation), automation zones (from physical equipment to market operations) and interoperability layers (from physical electrical connections to business interactions). Instance level models can be found in communication protocols and other standards for data exchange (e.g. SGMES) or configuration for simulations or equipment. We specifically modelled electrical equipment according to the data model of the simulation environment pandapower. 

\subsection{Auxiliary Layers for Interoperability and Analysis}
To relate concepts from different layers of abstractions, we define some auxiliary concepts in an intermediate layer between instance level and architectural models. Further, for specific analysis and tests, concepts are useful that are not explicitly stated in an existing model. We define those in an additional auxiliary layer. We abstractly modelled the configuration of firewalls, IP networks and communication flows for the use cases presented here. Since these abstract concepts are not part of any concrete specification, they are placed in such auxiliary layers.

\section{Automated Reasoning}
\label{sec:reasoning}
The analysis we perform on our data model as first use cases are safety and security testing.
To perform tests, we perform two kinds of reasoning on the A box knowledge in our model which we present in more detail in the following two subsections.
Shape validation with SHACL allows us on the one hand to verify the completeness of the assertional knowledge and on the other hand we formulate security and safety tests as shape constraints on the data graph.
The other kind of reasoning is making implicit assertional knowledge explicit through the evaluation of rules that add new statements to the knowledge base.
While our data model is openly extensible, following the paradigm of the semantic web, we reason under the closed world assumption, to be able to make definitive statements about the safety and security of particular instances of infrastructure. 

\subsection{Augmentation Reasoning}
To keep validation rules at a reasonable level of complexity, we augment data present in the A box through the evaluation of rules. In our prototypical implementation of our approach, we use SPARQL INSERT statements. Below is a simple rule, inserting the statement "a is of type b" if "a is of type c".

\begin{lstlisting}
    INSERT{
        ?a rdf:type ex:b .
    }
    WHERE{
        ?a rdf:type ex:c .
    }
\end{lstlisting}

Semantically, these statements are Horn clauses and we will use a prolog-style notation for these rules going forwards for compactness:

\begin{lstlisting}
    isB(a) :- isC(a)
\end{lstlisting}

We evaluate these rules naively. That means we continuously apply rules until no new statements are generated. We will replace this mechanism by a more stable reasoning regime in future work. 

\subsection{Shape Validation}
We use SHACL to make two kinds of statements of the data in our model: consistency and compliance to safety or security controls. To be able to reason accurately about assertional knowledge, we need to validate about the consistency of it. Under the closed world assumption, we deem all statements which are not included in our knowledge base to be false. Therefore we need to be sure that all required knowledge is explicitly stated. Once this is validated, we can run the validation of the security and safety controls, which are stated in the same framework as the consistency constraints.
SHACL defines shapes that concepts in the knowledge base need to conform to. A simple example is the necessity for a parent to have a child. A conforming database in RDF turtle syntax may look like this:

\begin{lstlisting}
    ex:p a ex:Parent .
    ex:p ex:hasChild ex:c .
\end{lstlisting}

The corresponding shape reads as follows:

\begin{lstlisting}
    ex:parentShape a sh:NodeShape ;
        sh:targetClass ex:Parent ;
        sh:property : [
            sh:path ex:hasChild ;
        ]
    .
\end{lstlisting}

\texttt{sh} being a shorthand for the SHACL namespace, \texttt{sh:targetClass} defining that all instances of the class \texttt{ex:Parent} will be matched against the shape and the \texttt{sh:property} expression defining that each such instance must have a path with the predicate \texttt{ex:hasChild}.

We will again use a more compact prolog-style notation for the examples, which is a headless statement:

\begin{lstlisting}
    :- isParent(x), childOf(x,y)
\end{lstlisting}

\section{Case Studies}
\label{sec:case-studies}
In this section we present two use cases for the proposed application. The scenarios are deliberately simple and abstracted from the data that would be available in the field to illustrate the approach as clearly as possible. Nonetheless, the use cases are inspired by real world incidents and thus point to the opportunity this approach might be.

For the prototypical implementation we use python implementations of the Semantic Web technologies introduced in the previous two sections; rdflib \cite{Krech2025-hn}(for in-memory processing of RDF graphs and execution of SPARQL queries) and pyshacl \cite{Sommer2025-zv} (for SHACL shape validation). Rules and data are kept in plain text files. We use turtle as serialization for RDF data.

\subsection{Network Separation}
\begin{figure}
    \centering
    \includegraphics[width=0.4\textwidth]{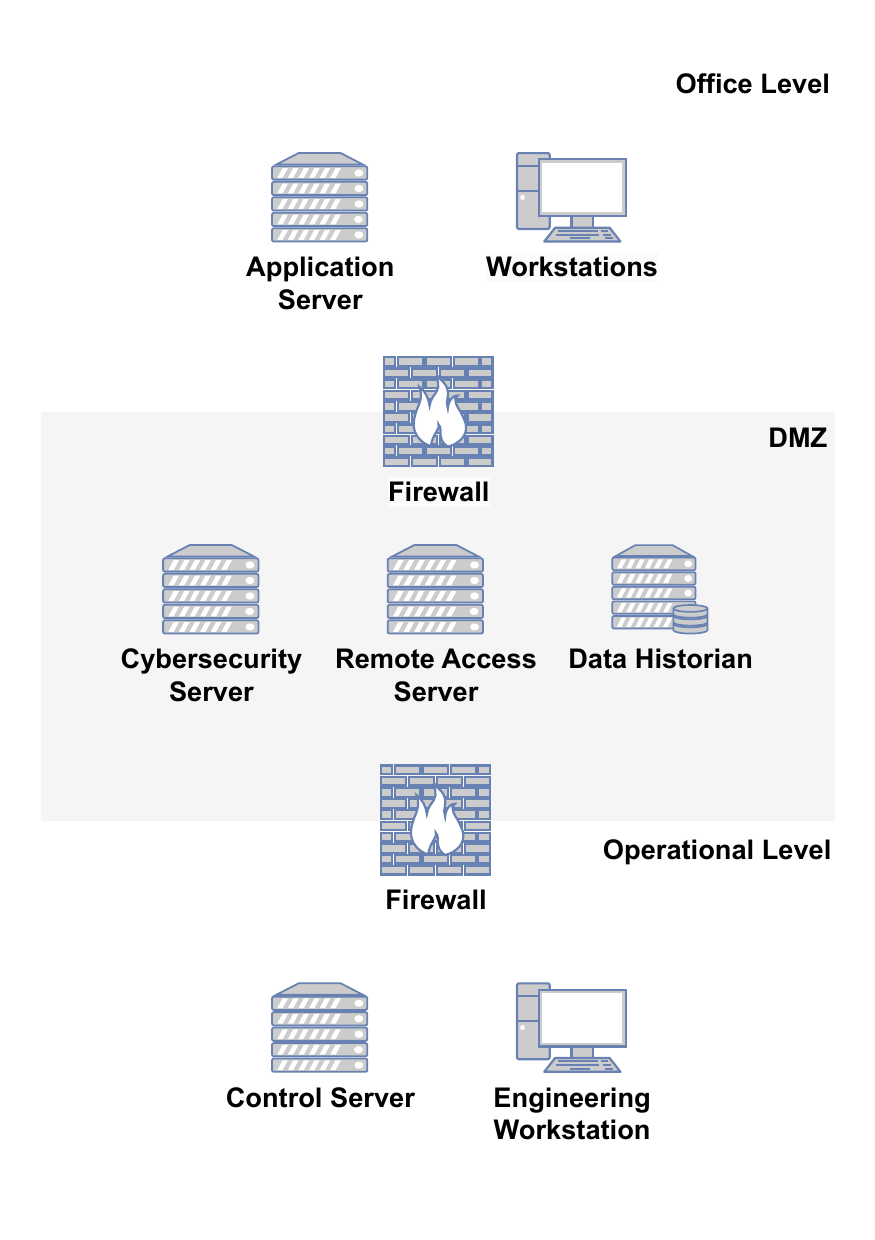}
    \caption{Simplified version of the defence-in-depth architecture example from \cite{Stouffer2023-fp} }
    \label{fig:nw-sep}
\end{figure}

Separating communication networks into zones with different levels of criticality (illustrated in figure \ref{fig:nw-sep}) is a standard aspect of the security posture of cyberphysical power systems. Networks are typically separated by firewalls which route traffic between zones based on rules. A misconfiguration of such a rule allowed malicious actors access to control infrastructure in the 2015 attacks on the Ukrainian power system \cite{Whitehead2017-eo}. 

Simplistically speaking, firewall rules designate source and destination address ranges between which traffic is permitted. For the illustrative implementation, we abstract away from explicit network addresses and instead define firewall rules as a binary relation between subnets (destination) and hosts or other subnets (source). This is a validation shape of a firewall rule:

\begin{lstlisting}
    :- isFWRule(x),
        source(x,y),
        destination(x,z),
        isSubnet(y),
        (isSubnet(z); isHost(z))
\end{lstlisting}

We also model the notion of two hosts being connected through a binary relation, which can be inferred by two hosts being connected to the same subnet or connected through a firewall. As an example, this is the augmentation rule that connects two hosts which are connected through a firewall.

\begin{lstlisting}
    connected(x,y) :- source(fw,sn1),
                      destination(fw,sn2),
                      hostInSubnet(x, sn1),
                      hostInSubnet(y,sn2)
\end{lstlisting}

After all augmentation rules are evaluated so often that no new knowledge is generated, we can validate the security control that states "Office Level hosts may not be connected to operational level hosts", which is formulated as such:

\begin{lstlisting}
    :- operationalHost(x),
       not connected(x,y),
       officeHost(y)
\end{lstlisting}

\subsection{Redundant Critical Communication}
\begin{figure}
    \centering
    \includegraphics[width=0.5\textwidth]{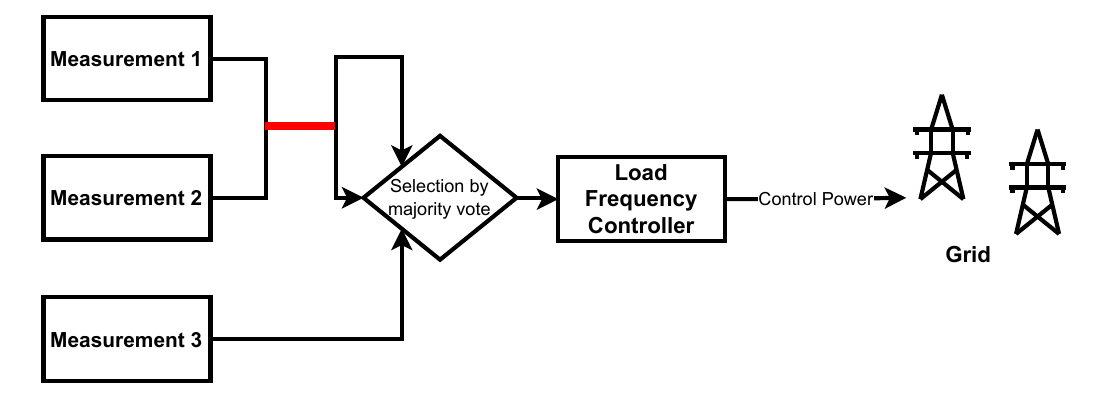}
    \caption{Simplified depiction of the data flow between the field station and the load frequency controller}
    \label{fig:freq-dev}
\end{figure}
In January 2019, continental Europe experienced a significant frequency deviation in its electrical transmission grid, leading to the disconnection of approximately 1.7 GW of load through the Industrial Interruptible Service by the French operator RTE. The frequency deviation was caused by a load frequency controller that was provided faulty measurements through channels that were thought to be redundant and hence fail-save. However, as illustrated in figure \ref{fig:freq-dev}, two of the three measurements were forwarded through the same substation and a drop of a telecommunication line between the source of the measurements and that substation led to the measurements to appear frozen at the load frequency controller \cite{ENTSO-E2019-oo}.

This misconfiguration could have been found if the infrastructure was modelled appropriately and the fail-safety was validated. For the illustrative implementation of this use cases, we only consider the architectural level of this scenario. We strive to link a closer representation of the physically deployed infrastructure to the architectural layer in a more mature iteration of our application. We model the scenario using concepts from SGAM: stations through which the measurements are routed are modeled as functional blocks, the measurements are modeled as payloads of information object flows. We can then derive which function blocks forward measurements:

\begin{lstlisting}
    forwards(f,m) :- payload(iof1, m),
                     source(iof1, f),
                     payload(iof2, m),
                     destination(iof2, f)
\end{lstlisting}

We can then infer the independence, i.e. fail safety of measurements by the abscence of function blocks forwarding both measurements through their incoming and outgoing information object flows:

\begin{lstlisting}
independent(m1,m2) :- forwards(f1, m1),
                      not forwards(f1, m2),
                      forwards(f2, m2),
                      not forwards(f2, m1)
\end{lstlisting}

This allows us to test the safety control that was violated in this scenario - that measurements feeding the load frequency controller should be independent:

\begin{lstlisting}
    :- loadFrequencyController(lfr),
       inputMeasurement(lfr, m1),
       inputMeasurement(lfr, m2),
       inputMeasurement(lfr, m3),
       independent(m1, m2),
       independent(m1, m3),
       independent(m2, m3)
\end{lstlisting}

\section{Conclusion and Future Work}
In this work we present an approach to security and safety testing of cyber-physical power systems grounded in formal logic and thorough modelling. We presented some first illustrative use cases, which were based on real incidents and underscore the potential of our approach. That is to automate the assessments that can be cleanly deduced from data available to operators and auditors and therefore improving the security and safety posture of the industry. 
However practical applicability of the approach hinges on a rich data model and strong semantics through rules and shapes.
We aim to make our work openly and freely available to gain applicability at scale, aiding in our effort to extend the model through additional external data models. At a sufficient scale, we will hopefully be able to conduct a thorough evaluation of the applicability of this approach to the real wold. We will also expand the system to serve more use cases, apart from the resilience use case described in \cite{Hacker2025-xg} and the safety and security use case described here. In the security realm we may proceed from testing individual security controls to automated threat modeling and attack graph generation. Apart from this domain, we see also potential in the planning of infrastructure and as a configuration mechanism for complex testing environments for cyber-physical power systems.
To serve those advanced aims, our system will potentially require a more mature reasoning system, to guarantee bounds on efficiency and termination.
\section{Acknowledgement}

\begin{wrapfigure}{r}{0.12\textwidth}
	\vspace{-\baselineskip}
	\vspace{-\baselineskip}
	\includegraphics[width=0.16\textwidth]{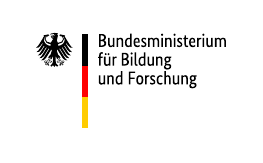}
	\vspace{-\baselineskip}
	\vspace{-\baselineskip}
	\vspace{-\baselineskip}
\end{wrapfigure}
This work has received funding from the  Federal  Ministry  of  Education and Research (BMBF) under project funding reference 03SF0694A.
\bibliographystyle{IEEEtran}
\bibliography{csr/csr}
\end{document}